\documentclass[acmsmall, screen]{acmart}

\usepackage{enumitem}
\usepackage{comment}
\usepackage{amsmath}
\usepackage{graphicx}
\usepackage{listings}
\usepackage{booktabs}
\usepackage{adjustbox}
\usepackage[skins]{tcolorbox}
\usepackage{subfigure}
\usepackage{fontawesome}
\usepackage{multirow}
\usepackage{wrapfig}
        
\AtBeginDocument{%
  \providecommand\BibTeX{{%
    \normalfont B\kern-0.5em{\scshape i\kern-0.25em b}\kern-0.8em\TeX}}}

\renewcommand\paragraph[1]{\subsubsection*{#1}}

\setcopyright{cc}
\setcctype{by}
\acmJournal{PACMSE}
\acmYear{2025}
\acmVolume{2}
\acmNumber{FSE}
\acmArticle{FSE013}
\acmMonth{7}
\acmDOI{10.1145/3715728}

\makeatletter
\newif\ifcolonfoundonthisline
\newcommand\JSONnumbervaluestyle{\color{blue!80!black}}
\newcommand\JSONstringvaluestyle{\color{red!80!black}}
\lstdefinestyle{json}{
  showstringspaces    = false,
  keywords            = {false,true},
  alsoletter          = 0123456789.,
  morestring          = [s]{"}{"},
  stringstyle         = \ifcolonfoundonthisline\JSONstringvaluestyle\fi,
  MoreSelectCharTable = \lst@DefSaveDef{`:}\colon@json{\processColon@json},
  basicstyle          = \footnotesize\ttfamily\linespread{0.5},
  keywordstyle        = \ttfamily\bfseries,
  numbers             = left,
  numberstyle         = \scriptsize,
  stepnumber          = 1,
  numbersep           = 6pt,
  showstringspaces    = false,
  breaklines          = true,
  frame               = lines,
  escapeinside        = {(*}{*)}
}
\newcommand\processColon@json{%
  \colon@json%
  \ifnum\lst@mode=\lst@Pmode%
    \global\colonfoundonthislinetrue%
  \fi
}
\lst@AddToHook{Output}{%
  \ifcolonfoundonthisline%
    \ifnum\lst@mode=\lst@Pmode%
      \def\lst@thestyle{\JSONnumbervaluestyle}%
    \fi
  \fi
  \lsthk@DetectKeywords%
}
\lst@AddToHook{EOL}{\global\colonfoundonthislinefalse}

\tcbset{
  my box2/.style={
    enhanced,
    colframe=#1!80,
    colback=#1!5,
  },
}
\newtcolorbox{finding}{
  my box2=black,
  boxrule=1pt,top=3pt,bottom=3pt,left=4pt,right=4pt
}

\newcommand{\Code}[1]{\begin{small}\texttt{#1}\end{small}}

\begin{document}

\title{Pinning Is Futile}
\subtitle{You Need More Than Local Dependency Versioning to Defend against Supply Chain Attacks}

\author{Hao He}
\orcid{0000-0001-8311-6559}
\affiliation{%
  \institution{Carnegie Mellon University}
  \city{Pittsburgh}
  \state{Pennsylvania}
  \country{USA}
}
\email{haohe@andrew.cmu.edu}

\author{Bogdan Vasilescu}
\orcid{0000-0003-4418-5783}
\affiliation{%
  \institution{Carnegie Mellon University}
  \city{Pittsburgh}
  \state{Pennsylvania}
  \country{USA}
}
\email{vasilescu@cmu.edu}

\author{Christian Kästner}
\orcid{0000-0002-4450-4572}
\affiliation{%
  \institution{Carnegie Mellon University}
  \city{Pittsburgh}
  \state{Pennsylvania}
  \country{USA}
}
\email{kaestner@cs.cmu.edu}

\begin{abstract}
Recent high-profile incidents in open-source software have greatly raised practitioner attention on software supply chain attacks. 
To guard against potential malicious package updates, security practitioners advocate \emph{pinning} dependency to specific versions rather than \emph{floating} in version ranges.
However, it remains controversial whether pinning carries a meaningful security benefit that outweighs the cost of maintaining outdated and possibly vulnerable dependencies.
In this paper, we quantify, through counterfactual analysis and simulations, the security and maintenance impact of version constraints in the npm ecosystem. 
By simulating dependency resolutions over historical time points, we find that pinning direct dependencies not only (as expected) increases the cost of maintaining vulnerable and outdated dependencies, but also (surprisingly) even increases the risk of exposure to malicious package updates in larger dependency graphs due to the specifics of npm's dependency resolution mechanism. 
Finally, we explore collective pinning strategies to secure the ecosystem against supply chain attacks, suggesting specific changes to npm to enable such interventions.
Our study provides guidance for practitioners and tool designers to manage their supply chains more securely.
\end{abstract}



\maketitle

\begin{table}
\renewcommand{\arraystretch}{1.1}
\definecolor{darkred}{rgb}{0.9, 0.0, 0.0}
\definecolor{darkgreen}{rgb}{0.0, 0.8, 0.0}
    \small
    \centering
    \caption{The commonly argued trade-offs (\emph{\color{darkgreen}\faToggleDown} benefits and \emph{\color{darkred}\faToggleUp} drawbacks) between pinning and floating}
    \begin{tabular}{l@{~}p{9.8cm}ll}
    \toprule
        \multicolumn{2}{l}{The Trade-Offs (\faLock~~Security and {\faCogs}~~Maintenance)} & Floating & Pinning  \\
    \midrule
        \faLock & \textbf{Attack Surface for Malicious Package Updates}~\cite[e.g.,][]{openssf-pinning, facts-vs-feelings, DBLP:journals/ieeesp/ZahanKHSW23, DBLP:conf/sp/LadisaPMB23} & \emph{\color{darkred}\faToggleUp}~Larger & \emph{\color{darkgreen}\faToggleDown}~Smaller \\
        \faCogs & Effort to Handle Breaking Changes~\cite[e.g.,][]{pin-your-npm-yarn-dependencies, stop-version-range, pin-exact-dep-versions, 
        DBLP:journals/tse/JafariCAST22, DBLP:journals/jss/RaemaekersDV17, DBLP:journals/tosem/BogartKHT21} & \emph{\color{darkred}\faToggleUp}~More & \emph{\color{darkgreen}\faToggleDown}~Less  \\[.5em]
        \faLock & \textbf{Attack Surface from Security Vulnerabilities}~\cite[e.g.,][]{floating-discussion, should-i-pin-python-deps, DBLP:journals/tse/JafariCAST22, DBLP:journals/ese/ChinthanetKMIIM21, DBLP:journals/tse/0038SP00C0023} & \emph{\color{darkgreen}\faToggleDown}~Smaller & \emph{\color{darkred}\faToggleUp}~Larger \\
        \faCogs & Effort to Update Outdated Dependencies~\cite[e.g.,][]{
        DBLP:journals/tse/JafariCAST22} &  \emph{\color{darkgreen}\faToggleDown}~Less & \emph{\color{darkred}\faToggleUp}~More \\
        \faCogs & Effort to Resolve Dependency Conflicts~\cite[e.g.,][]{floating-discussion} 
         & \emph{\color{darkgreen}\faToggleDown}~Less & \emph{\color{darkred}\faToggleUp}~More \\
    \bottomrule
    \end{tabular}
    \label{tab:trade-offs}
\end{table}

\section{Introduction}
\label{sec:introduction}

Open-source components have become a critical part of software supply chains in today's software development. 
However, recent high-profile security incidents show that integrating open-source components exposes a software project to two major types of security risks:
\begin{itemize}[leftmargin=20pt]
    \item \textbf{Security Vulnerabilities:} An open-source component, especially outdated versions, may contain \textit{accidentally} introduced vulnerabilities or weaknesses, which may be exploitable by a malicious actor in certain cases (e.g., the Log4Shell vulnerability~\cite{Log4Shell}). 
    \item \textbf{Malicious Package Updates:} An open-source component may be \textit{intentionally} compromised by an attacker to release a malicious update and exploit downstream applications (e.g., the \Code{node-ipc}~\cite{node-ipc} and \Code{xz} incidents~\cite{xz-utils}).
    This scenario is often referred to, in a broader sense, as \emph{software supply chain attacks}~\cite{DBLP:conf/dimva/OhmPS020, DBLP:conf/sp/LadisaPMB23}.
\end{itemize}

In most ecosystems, software projects need to specify \emph{version constraints} for their dependencies on open-source components---\textit{these constraints influence their exposure to security vulnerabilities and malicious package updates}. 
For example, npm packages are expected to follow semantic versioning~\cite{semver}, using minor version updates to indicate non-breaking releases.
Then, downstream developers can use \emph{floating} version constraints (e.g., ``\Code{\^{}1.2.0}'' matches \Code{1.2.0} $\le$ version $<$ \Code{2.0.0}) to allow their project to automatically install the most recent, patched but also backward compatible version of a dependency at the time of installation.
Through this mechanism, a security patch fixing a vulnerability (e.g., \Code{1.2.1} fixing \Code{1.2.0}) can be immediately propagated to a downstream project as long as it uses floating version constraints for that dependency.

However, the same floating mechanism can also lead to the immediate propagation of malicious package updates in case a dependency gets compromised and releases a malicious version as a new patch release (e.g., the \Code{node-ipc} incident~\cite{node-ipc}).
Even if malicious updates are less common, they are arguably more concerning than security vulnerabilities because many security vulnerabilities are hard to exploit~\cite{DBLP:conf/codaspy/YounisMAR16, DBLP:journals/corr/abs-1908-04856} whereas the injection of malicious code can cause immediate catastrophic harm and is difficult to defend against.
To prevent exposure to malicious package updates, it seems necessary to \emph{pin} dependency versions (e.g., require exactly version \Code{1.2.0}). However, this strategy comes with its own costs.
To receive future updates, developers have to manually update the dependency to a newer release, ideally after auditing the changes. 
Failure to do so can leave security vulnerabilities to hackers even when the security patch has been available for a long time~\cite{DBLP:conf/msr/DecanMC18}.

The choice between \emph{pinning} and \emph{floating} not only represents a \emph{trade-off} among competing security priorities but also interacts with other dependency management concerns.
Specifically, the above discussion also applies to breaking changes (that are not malicious) and outdated dependencies (a form of technical debt).
What's more, modern software supply chains form complex networks in which each node (i.e., project) can only make local decisions about its direct dependencies~\cite{DBLP:journals/ese/DecanMG19}, but all pinning and floating decisions an upstream project makes have implications, \textit{both directly and transitively,} for all its downstream projects.
For example, a floating version constraint leaves an attack surface for all downstream projects, while pinning puts downstream projects at a higher risk of encountering dependency conflicts~\cite{DBLP:conf/sigsoft/WangWLWWYYZC18, DBLP:conf/icse/PinckneyCGBCG23}.
We summarize the commonly argued benefits and drawbacks of pinning and floating in Table~\ref{tab:trade-offs}.

\begin{figure}[t]
    \centering
    \subfigure[The distribution of version constraint types.]{\includegraphics[width=0.49\linewidth]{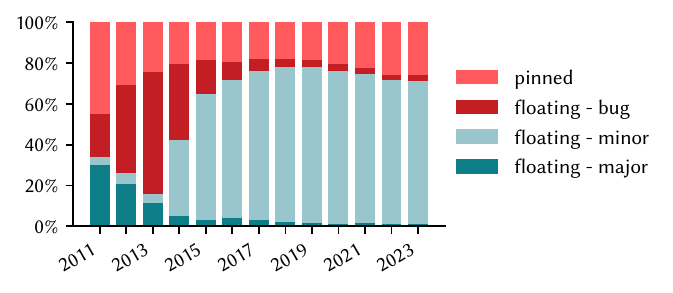}\label{fig:versioning}}
    \subfigure[The distribution of project-level versioning strategies.]{\includegraphics[width=0.46\linewidth]{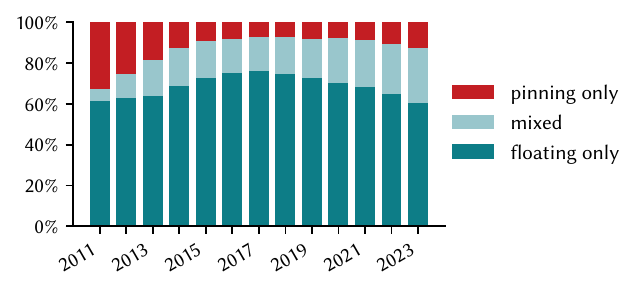}\label{fig:versioning-strategy}}
    \caption{
    The distribution of version constraint types and project-level versioning strategies in each year in the dataset we use for answering RQ1 (see Section~\ref{sec:data-collection} for details) - we observe that after years of increasing floating adoption, since about 2020 the trend reverses with a shift toward pinning. In Figure~\ref{fig:versioning}, we ignore non-floating/pinning version constraints as they only occupy 0.83\% of version constraints in our dataset.
    }
    \label{fig:version-dist}
\end{figure}

Importantly, the concern about \emph{malicious package updates} is fairly recent and reached broad awareness through high-profile incidents, such as \textit{SolarWinds} in 2020 \cite{solar-wind}, \Code{node-ipc} in 2022~\cite{node-ipc} and the \Code{xz} attack targeted at \Code{ssh} in 2024~\cite{xz-utils}---such recent attention may shift trade-off considerations.
Prior to this recent concern, 
there have been many studies and discussions about the trade-off between semantic versioning and breaking changes~\cite[e.g.,][]{DBLP:journals/jss/RaemaekersDV17, DBLP:journals/tosem/BogartKHT21, DBLP:journals/tse/JafariCAST22}.
In a survey conducted in 2020, the majority of practitioners in the npm ecosystem were in favor of (semantic-versioning-based) floating~\cite{DBLP:journals/tse/JafariCAST22}.
It is argued that pinning accumulates technical debt in the form of outdated dependencies~\cite{i-have-misgivings-about-all-these-version-pinning, DBLP:journals/tse/JafariCAST22}, prevents security fixes from propagating downstream~\cite{DBLP:journals/ese/ChinthanetKMIIM21, DBLP:journals/tse/0038SP00C0023}, and increases the likelihood of dependency conflicts (or bloat in npm)~\cite{floating-discussion, should-i-pin-python-deps}.
However, recent high-profile malicious package update incidents provide a new argument in favor of pinning, as pinning reduces the attack surface for malicious package updates.
Given this argument, together with prior arguments that pinning increases application stability and reproducibility~\cite{pin-your-npm-yarn-dependencies, stop-version-range, pin-exact-dep-versions}, 
many practitioners now argue in favor of pinning for security sensitive applications --
for example, a survey of 17 domain experts and 134 software developers by \citet{DBLP:conf/sp/LadisaPMB23} ranks pinning as a top-3 safeguard against supply chain attacks in terms of utility/cost ratio; the OpenSSF Scorecard project~\cite{openssf-pinning, DBLP:journals/ieeesp/ZahanKHSW23} included ``pinned dependencies'' as one of its 18 security best practices.
This discussion clearly reaches practitioners---as we show in Figure~\ref{fig:version-dist} (and explain in more depth later in Section~\ref{sec:data-collection}), practitioners in npm have gradually shifted toward pinning more dependencies since 2020.

Overall, controversies remain on whether the benefits of pinning outweigh its drawbacks, as manifested by online discussions~\cite{facts-vs-feelings, renovate-discussion, pinning-only-work} and evidenced by the contradicting results from the collected practitioner opinions in prior studies \cite{DBLP:journals/tse/JafariCAST22, DBLP:conf/sp/LadisaPMB23}.
However, to the best of our knowledge,  \textit{there is no quantitative evidence so far about the actual impact of pinning versus floating dependencies on the security and maintenance trade-offs in Table~\ref{tab:trade-offs}}.
To gather quantitative evidence about pinning and floating, we perform a repository mining study on the npm ecosystem, the most frequently discussed ecosystem in the literature.
Our study highlights an important complication that makes the discussion around pinning so difficult: Since a project can only control the version constraints of its direct dependencies, the actual impact of pinning as a local intervention is complicated and unclear, contingent upon both the decisions made upstream and the behaviors of dependency management tools. 
To measure this impact holistically from empirical data, we raise the following research question:
\begin{itemize}[leftmargin=20pt]
    \item \textbf{RQ1:} (\emph{Impact of Pinning}) How does the individual choice of pinning direct dependencies affect the overall security risks and maintenance costs in one's own dependency graph?
\end{itemize}
To answer this research question, we collect a large sample of npm packages and GitHub repositories using npm.
Then, we use an experimental time-traveling feature of the npm dependency resolver to simulate the evolution of their dependency graphs over a 12-month period.
We estimate the impact of pinning on a set of five security and maintenance-related metrics using \emph{fixed-effect panel regression}~\cite{bruderl2015fixed, hanck2021introduction}.
A key (and counter-intuitive) finding from our model is that for the purpose of reducing the attack surface of malicious updates, \textit{pinning is futile and can be even harmful} for projects with $\ge$498 direct and transitive dependencies (affecting 26\% of GitHub repositories in our dataset). That is, pinning is ineffective for one of its main goals, all while creating substantial costs for managing outdated, vulnerable, and conflicting dependencies. 

Inspired by this finding, we argue that \textit{ecosystem-level interventions} (e.g., investing resources to secure critical open-source projects) are more effective than (possibly futile) local interventions of pinning direct dependencies to improve the overall resilience of an open-source ecosystem against malicious package updates.
To empirically test this, we explore the second research question:
\begin{itemize}[leftmargin=20pt]
    \item \textbf{RQ2:} (\emph{Ecosystem-Level Interventions}) How effective is coordinating pinning among selected upstream packages in reducing the risk of malicious package updates ecosystem-wide? 
\end{itemize}
We answer this research question by simulating coordinated efforts in the entire npm dependency network.
We formulate attack models and defense strategies in the dependency network, define heuristics to intervene in selected upstream packages, and measure the overall risk of malicious package updates after the intervention.
Our findings indicate that it is possible to reduce the average spread of malicious package updates in the npm dependency network by nearly 30\% if we pin in 100 intentionally selected important packages.
However, as we will quantify, this intervention could be substantially more effective (up to 75\%) if npm implements a new \textit{transitive pinning} feature to allow packages to pass their entire pinned dependency graphs downstream.

\textsf{\emph{Summary of Contributions.}} Our paper contributes novel quantitative evidence on the security and maintenance impact of version constraints in a packaging ecosystem, especially in light of recent concerns about malicious package updates.
Our findings align with the opinions of npm developers~\cite{DBLP:journals/tse/JafariCAST22} but contradict the views of security practitioners~\cite{DBLP:conf/sp/LadisaPMB23, DBLP:journals/ieeesp/ZahanKHSW23}, suggesting that pinning direct dependencies is futile (and can be even counter-productive) as protection against malicious package updates.
In addition, we contribute evidence that a new ecosystem-wide intervention, along with tooling changes and ecosystem-wide allocation of collectively pooled resources, can be an effective intervention against malicious package updates.
Our findings lead to concrete advice for developers and ecosystem policy makers to improve the state of software supply chain security (Section~\ref{sec:discussion}). 

\section{Background and Related Work}
\label{sec:background}

\paragraph{Dependency Management in npm}

\begin{figure}[b]
    \scriptsize
    \centering
\begin{lstlisting}[style=json,numbers=none]
{ "name": "postcss-loader", "version": "7.3.3",
  "dependencies": {
    "webpack": ">=5.0.0"     (*\color{red}$\leftarrow$~\textsf{\textbf{floating - major}}*)
    "cosmiconfig": "^8.3.5", (*\color{red}$\leftarrow$~\textsf{\textbf{floating - minor: $\ge$8.3.5, $<$9.0.0}}*)
    "jiti": "~1.21.0",       (*\color{red}$\leftarrow$~\textsf{\textbf{floating - patch: $\ge$1.21.0, $<$1.22.0}}*)
    "semver": "7.5.4"        (*\color{red}$\leftarrow$~\textsf{\textbf{pinned}}*)
    "postcss": "^7.0.0 || ^8.0.1", (*\color{red}$\leftarrow$~\textsf{\textbf{other}}*)
    "lodash": "git+ssh://git@github.com:lodash/lodash.git#v4.17", (*\color{red}$\leftarrow$~\textsf{\textbf{other}}*)
  }, "devDependencies": {...}, ... }
\end{lstlisting}
\caption{
An example \Code{package.json} file. 
We modified version constraints in the original file to illustrate different types of version constraints defined in Section~\ref{sec:method-rq1}. Other details irrelevant to this paper are omitted.}
\label{fig:example-pkgjson}
\end{figure}

npm is the most widely used dependency manager for  JavaScript and TypeScript~\cite{npm}.
It consists of a package registry that hosts open-source components (called \emph{packages} in npm) and a command-line tool that helps projects reuse packages from the package registry.  
As of September 2024, the npm registry is the largest package registry, hosting more than 3 million packages and 25 million releases~\cite{module-counts}.
The npm command-line tool reads direct dependencies from a configuration file named \Code{package.json}.
In the \Code{package.json} file (see Figure~\ref{fig:example-pkgjson} for an example), developers specify package names as the \emph{dependencies} of the project with a \emph{version constraint} for each dependency.
In addition, npm distinguishes different kinds of dependencies---usually, developers focus more on \emph{production dependencies}~\cite{DBLP:conf/kbse/LatendresseMCS22}, i.e., dependencies that will be included in the production build (for applications) or passed downstream (for packages), and are less concerned about security vulnerabilities if the dependencies are not used in production~\cite{DBLP:conf/esem/PashchenkoPPSM18, DBLP:conf/kbse/LatendresseMCS22}.

Given a \Code{package.json} file, npm will resolve a \textit{dependency graph} including not only dependencies specified in the \Code{package.json} file, but also all transitive dependencies required by dependencies.
The npm ecosystem is known as particularly fragmented with many tiny packages and high levels of reuse~\cite{DBLP:conf/sigsoft/AbdalkareemNWMS17, DBLP:journals/ese/DecanMG19}; a project commonly has orders of magnitude more transitive dependencies than direct ones~\cite{DBLP:journals/ese/DecanMG19}.
Dependency resolution can be complicated when multiple dependencies depend on the same third dependency but with conflicting version constraints; this is known as the \textit{diamond problem} or \textit{dependency hell} (see Figure~\ref{fig:rq2-example}). 
While it is highly problematic in other ecosystems~\cite{fan2020escaping,mancinelli2006managing,DBLP:conf/wcre/AbateCGZ20,DBLP:journals/tosem/BogartKHT21}, npm can install multiple distinct versions for the same package, but it will generally try to produce a dependency graph with fewer duplicate versions (i.e., deduplication~\cite{npm-install}). 
Optimal deduplication is NP-Complete~\cite{DBLP:conf/wcre/AbateCGZ20}, so npm uses a variety of heuristics to search for a ``good'' dependency graph satisfying all the version constraints while having fewer duplicate versions~\cite{npm-install}. 

Importantly, \emph{dependency graphs are volatile}. 
Since npm always installs the latest version inside a floating versioning constraint, the same \Code{package.json} may result in different graphs as new package updates are released~\cite{DBLP:conf/icse/LiuCF00022}.
With floating, projects can keep dependencies up-to-date automatically, whereas developers often do not update pinned dependencies timely, even for those with known security vulnerabilities~\cite{DBLP:conf/icse/CoxBEV15, DBLP:journals/ese/KulaGOII18, DBLP:journals/tse/HeHZZ23}.

A key challenge to floating dependencies is \textit{breaking changes}: When an update changes expected behavior, the application can break without any action by its developer. Developers then need to react and either adapt their own code or downgrade the dependency~\cite{DBLP:journals/tosem/BogartKHT21}.
To control breaking changes, npm advocates for \emph{semantic versioning}~\cite{npm-semver}: A release should increment its major version number (e.g., from \Code{1.2.1} to \Code{2.0.0}) if it contains breaking changes, minor version number (e.g., from \Code{1.2.1} to \Code{1.3.0}) if it contains new features, and patch version number (e.g., from \Code{1.2.1} to \Code{1.2.2}) if it contains only bug fixes. If packages follow those conventions, automatic minor and patch updates should be safe. 
To support this practice, npm provides specific patterns to declare version ranges accepting minor (e.g., \Code{\^{}8.3.5}) and patch updates (e.g., \Code{\~{}1.21.0}); it also adopts floating versions for minor updates by default when a developer adds a dependency with  \Code{npm install}.
Although previous research has shown that 
packages may fail to obey semantic versioning~\cite{DBLP:journals/jss/RaemaekersDV17, semantic-versioning-sucks, DBLP:conf/kbse/ZhangLXCF0022, DBLP:conf/issta/JayasuriyaT0OB23} and developers may turn to pinning instead if they have lost trust in its reliability~\cite{stop-version-range, semantic-versioning-sucks},
 semantic-versioning-based floating is still widely accepted and used by npm developers~\cite{DBLP:conf/msr/0001PSTB19, DBLP:journals/tse/DecanM21, DBLP:conf/msr/PinckneyCGB23, DBLP:journals/tse/JafariCAST22}.
In fact, without discussion of supply chain attacks, the survey by~\citet{DBLP:journals/tse/JafariCAST22} shows that npm developers generally consider pinned dependencies as a ``smell,'' indicating a compromise that will accumulate technical debt and bring negative consequences in the long term.

Finally, it is worth noting that the above discussion applies not only to floating direct dependencies but also to floating transitive dependencies even if the developer pins all their direct dependencies.
Developers usually have no control over how their dependencies float or pin transitive dependencies.
As a workaround, npm (as many other ecosystems) provides functionality to \textit{freeze} and store a project's entire dependency graph with a lock file (\Code{package-lock.json}~\cite{npmpkglock}). 
A lock file ensures the reproducibility of npm builds, but
it has two limitations: (1)~it is restricted to applications (the root of the dependency graph) whereas components used as dependencies cannot pass their lock decisions to downstream applications; it is hence not very useful for packages; (2)~it is not designed (hence very difficult) to be manually maintained, so floating version constraints in \Code{package.json} will still cause automatic updates every time a lock file is generated.

\paragraph{Software Supply Chain Security}

The term \textit{software supply chain} is an umbrella term referring to all the software components, processes, and tools that a software project relies on~\cite{software-supply-chain-guide}.
For a project using npm, all packages in its dependency graph are part of its supply chain.
Most software projects nowadays are built upon a collaboratively developed open-source software supply chain~\cite{eghbal2016roads}; the 3~million packages in the npm package registry are a prominent example.

Software supply chains are fragile.
Such fragility comes from the collaborative and distributed nature of open-source development: Projects are often built with a large number of open-source components, each of them maintained by different groups of developers~\cite{schueller2024modeling}.
Meanwhile, popular components are directly or transitively used by a large number of downstream applications.
Consequently, any single weakness in the software supply chain could have a huge impact on the entire ecosystem~\cite{DBLP:journals/ese/DecanMG19, DBLP:conf/uss/ZimmermannSTP19}, creating a large attack vector for malicious actors and making the software industry increasingly concerned with software supply chain security~\cite{DBLP:journals/usenix-login/GeerTM20,DBLP:conf/sp/LadisaPMB23}.

In recent years, a significant amount of practitioner attention has been drawn to the rising threat of \emph{software supply chain attacks}~\cite{DBLP:conf/dimva/OhmPS020, DBLP:conf/sp/LadisaPMB23}. 
Although the attackers may attempt to trick malware installation in other ways (e.g., typosquatting~\cite{DBLP:conf/eurosp/VuPMPS20}), the most concerning incidents of software supply chain attacks are \emph{malicious package updates}, in which malicious actors \textit{intentionally} compromise an open-source component to inject malware into its (often many) downstream dependents.
While not every security vulnerability is exploitable in practice (leading developers to treat them as false positives)~\cite{DBLP:conf/codaspy/YounisMAR16, DBLP:journals/corr/abs-1908-04856, mirhosseini2017can}, malicious package updates can immediately harm the infected downstream dependents~\cite{martinez2021software}. 
They may also be well obfuscated and often very difficult to detect~\cite{xz-utils}.
Their high catastrophic potential and lack of controllability amplify the signal of real incidents and elevate public risk perception (e.g., in a way similar to nuclear hazards~\cite{slovic1987perception}).
As a result, malicious package updates are considered one of the primary threats in today's threat landscape~\cite{enisa-report, whitehouse-order}.
 
As our previous discussion of the pinning versus floating debate illustrates, many dependency management practices have security implications and involve trade-offs between conflicting priorities~\cite{DBLP:conf/ccs/PashchenkoVM20}.
Previous studies have investigated various practices, such as dependency adoption, updates, abandonment, migration, and the use of dependency management tools~\cite[e.g.,][]{DBLP:journals/ese/KulaGOII18, DBLP:conf/sigsoft/VargasATBG20, DBLP:conf/sigsoft/HeHGZ21, DBLP:conf/sigsoft/MillerKV23, DBLP:journals/tse/HeHZZ23}.
It is worth noting, however, that most prior research in the software engineering community is centered around the mitigation of security vulnerabilities.
On the other hand, the security community has recently consolidated best practices to defend against supply chain attacks, such as the OpenSSF Scorecard project~\cite{DBLP:journals/ieeesp/ZahanKHSW23} and the survey by \citet{DBLP:conf/sp/LadisaPMB23} introduced in Section~\ref{sec:introduction}.
However, these (supposed) security best practices may interact with existing \emph{software engineering} trade-offs, among which the pinning versus floating problem is particularly complicated (Table~\ref{tab:trade-offs});  the strongly held convictions from both sides lead to a considerable amount of controversy among practitioners~\cite{facts-vs-feelings, renovate-discussion, pinning-only-work}.

\paragraph{Novelty of Our Research.} 
There have been many prior studies on the use of version constraints~\cite{DBLP:conf/msr/0001PSTB19, DBLP:journals/tse/DecanM21, DBLP:conf/msr/PinckneyCGB23, DBLP:conf/sp/LadisaPMB23} and the trade-offs  around semantic versioning with regards to maintenance effort, breaking changes, and security vulnerabilities~\cite{DBLP:journals/tse/JafariCAST22, DBLP:journals/tosem/BogartKHT21,DBLP:journals/jss/RaemaekersDV17}.
However, the new threat of \textit{malicious package updates} complicates the previously discussed trade-offs, as \emph{malicious actors intentionally break semantic versioning}, often disguising their attacks as patch releases for maximum propagation~\cite{node-ipc}.
The changed threat landscape and the complexity of this pinning versus floating problem eventually lead to controversies and divergence in practitioner opinions~\cite{facts-vs-feelings, renovate-discussion, pinning-only-work, DBLP:conf/sp/LadisaPMB23, DBLP:journals/tse/JafariCAST22}.
Prior studies are either collecting these opinions~\cite{DBLP:journals/tosem/BogartKHT21, DBLP:journals/tse/JafariCAST22, DBLP:conf/sp/LadisaPMB23} or aggregating descriptive statistics from software repositories~\cite{DBLP:conf/msr/0001PSTB19, DBLP:conf/msr/PinckneyCGB23, DBLP:journals/tse/DecanM21, DBLP:journals/jss/RaemaekersDV17}; we are not aware of any prior study that attempted to quantify the impact of version constraints using statistical modeling methods.
To shed light on resolving the controversy, we believe it is vital to conduct data-driven research and quantitatively measure the actual impact of version constraints while taking all the previously discussed trade-offs into consideration (Table~\ref{tab:trade-offs}).
This forms the main motivation of our study.

\section{RQ1: How does the individual choice of pinning direct dependencies affect the overall security risks and maintenance costs in one's own dependency graph?}

As we discussed above, dependency pinning is often acknowledged as a security best practice to reduce the attack surface for malicious package updates~\cite{DBLP:conf/sp/LadisaPMB23, DBLP:journals/ieeesp/ZahanKHSW23}, but controversies remain on whether the security benefits of pinning outweigh its maintenance drawbacks~\cite{facts-vs-feelings, renovate-discussion, pinning-only-work, DBLP:conf/sp/LadisaPMB23, DBLP:journals/tse/JafariCAST22}.
Importantly, a project only has control over the version constraints of its direct dependencies, but the impact of pinning direct dependencies is complicated and unclear, contingent upon upstream practices and dependency management tools.
Hence, the goal of RQ1 is to empirically measure the actual security and maintenance impact of this intervention (i.e., pinning direct dependencies).

\subsection{Data Collection}
\label{sec:data-collection}

We aim to build a balanced dataset of packages (e.g., libraries, frameworks) and applications (e.g., websites, tools), as both need to manage dependencies but may differ in important ways.
Thus, we use two data sources: (1) \emph{npm packages}, (2) \emph{GitHub repositories using npm} that are not npm packages.
Note that neither data source is a clear representation of libraries or applications. 
Some npm packages may be applications and some GitHub repositories may be libraries that have not been uploaded to the npm registry. 
Still, combining the two should produce a stratified sample including both representation of libraries and applications in the wild.

For the first data source, npm packages, we curate a large dataset of popular and non-obsolete npm packages. We avoid focusing only on the few most popular packages, as they are not necessarily representative of common development practices in npm (they tend to occupy an upstream position with no dependencies or a very small dependency graph). 
We use the following data collection process:
First, we use npm-follower~\cite{DBLP:conf/sigsoft/PinckneyCG023} (September 2023 dump) to obtain a complete list of npm packages and releases at that time.
Then, we collect year-long download statistics as a popularity measure (between September 06, 2022, and September 05, 2023) using the npm API. For each package, we also identify its corresponding GitHub repository from npm metadata.
Next, using this dataset, we select the top 5\% most downloaded packages (21,218+ downloads) that have links to a GitHub repository and discard the bottom 25\% of these packages with the lowest number of stars and issues in GitHub (specifically, less than 10 stars and 8 issues), to ensure that the remaining are actively maintained open-source packages.
To further ensure they are not obsolete, we discard packages that have no release in the last two years at the time of dataset cutoff (September 2023).
We discard packages without any production dependencies in the latest release and end up with 41,516 npm packages.

For the second data source, we curate a dataset of GitHub repositories using npm, but are not npm packages, from World of Code~\cite{DBLP:journals/ese/MaDBAVTKZM21} (version V).
First, we select the JavaScript and TypeScript projects with top 1\% stars (i.e., \# stars $\ge$ 14) as a rough popularity threshold from its deduplicated GitHub metadata~\cite{DBLP:conf/msr/SpinellisKM20}.
To filter out small (and likely toy) projects, we discard projects with the bottom 25\% of commits and active months (specifically, $\le$84 commits and $\le$13 active months).
Then, we filter out projects that have been inactive (no commit) in the last two years at the time of dataset cutoff (September 2023).
To find GitHub repositories using npm, we retain projects with a \Code{package.json} file in their root folder. 
Finally, we discard repositories without any production dependencies and repositories whose \Code{name} field in \Code{package.json} matches a package name in npm---thus removing the ones more likely to be npm packages themselves.
This results in a total of 12,197 GitHub repositories.

As the two data sources are heavily imbalanced and the simulations (see Section~\ref{sec:method-rq1}) are computationally expensive, we downsample the two data sources to 10,000 npm packages and 10,000 GitHub repositories, respectively (uniform random sampling each).
We chose 10,000 each because it is computationally feasible for our simulation (in our case, two weeks of computation) and offers very high statistical generalizability.
In the remainder of RQ1, we will refer to the 10,000 npm \textit{packages} and the 10,000 GitHub \textit{repositories} collectively as \emph{projects}.

\begin{table}[t]
    \scriptsize
    \centering
    \caption{Descriptive statistics of the studied npm packages and GitHub repositories (excluding npm packages). The number of transitive dependencies is an estimate using successful dependency resolutions (Section~\ref{sec:method-rq1}).}
    \begin{tabular}{lrrrrrrrr}
      \toprule
         & \multicolumn{4}{c}{\textbf{npm packages}} & \multicolumn{4}{c}{\textbf{GitHub repositories (excl. npm packages)}} \\
         & Min & Median & Mean & Max & Min & Median & Mean & Max\\
      \midrule
       \# Annual Downloads    & 21,234  & 234K & 24.7M & 516M & \multicolumn{4}{c}{N/A}\\ 
       \# Stars         & 8 & 345 & 6,431.80 &  220K & 14 & 62 & 694.72 & 330K\\
       \# Issues        & 10 & 569 & 6338.55  & 87,258 & 0 & 101 & 652.62 & 75,561\\
       \# Commits       & 1 & 1,728 & 30,064.11 & 10.1M & 84 & 632 & 6215.72 & 7.17M\\
       \# Active Months & 1 & 48 & 60.7 & 324 & 13 & 36 & 44.26 & 323\\ 
       \# Direct Dependencies (Production)  & 1 & 3 & 5.78 & 283 & 1 & 11 & 16.83 & 492 \\
       \# Direct Dependencies (All) & 1 & 10 & 14.58 & 283 & 1 & 23 & 32.03 & 502 \\
       \# Transitive Dependencies (Production) & 0 & 11 & 65.55 & 2,320 & 0 & 150 & 391.51 & 3,722\\
       \# Transitive Dependencies (All) & 0 & 253 & 414.74 & 3,141 & 0 & 848 & 893.18 & 4,013\\
      \bottomrule
    \end{tabular}
    \label{tab:dataset}
\end{table}

\paragraph{Dataset Overview.}
We summarize the descriptive statistics from both data sources in Table~\ref{tab:dataset}.
Most of the projects in our dataset are popular and active.
As expected, GitHub repositories tend to have a much larger dependency graph than npm packages.
A median GitHub project has 11 direct dependencies and 150 transitive dependencies in production (23 direct and 848 transitive if we include development dependencies).
This confirms our intuition that many projects have orders of magnitude higher numbers of transitive dependencies compared with direct ones. 

In Figure~\ref{fig:version-dist}, we provide an overview of the use of version constraints in our dataset,
which largely mirrors findings of prior studies~\cite{DBLP:conf/msr/0001PSTB19, DBLP:journals/tse/DecanM21, DBLP:conf/msr/PinckneyCGB23}, but shows a previously not observed change in trend in more recent data, where developers start to pin more dependencies since around 2020.
Specifically, we aggregate dependency declarations of each project in each year, looking back at past revisions (releases for npm packages and commits for GitHub repositories).  
To reduce noise introduced by projects with a massive number of revisions, we sample one revision for each project in each year, as also commonly done in prior studies~\cite[e.g.,][]{DBLP:conf/msr/PinckneyCGB23, DBLP:conf/kbse/XuHGZ23}.
For each revision, we parse the version constraints of all its production dependencies in its \Code{package.json} file, according to the npm semver syntax~\cite{npm-semver}, and categorize the version constraints into five different types: (1) \emph{floating - major}, (2) \emph{floating - minor}, (3) \emph{floating - patch}, (4) \emph{pinned}, and (5) \emph{other}. 
Examples of these different version constraint categories can be found in Figure~\ref{fig:example-pkgjson}.
Overall, we find that semantic-versioning-based floating is used the most in the npm ecosystem but  pinning is still the second most popular choice, with a rising popularity since around 2020, even though npm does not advocate pinning~\cite{npm-semver}.
This aligns with our intuition that the rising concern of malicious package updates triggers changes in practitioner practices.
To further inform our study design, we examine how individual projects demonstrate preferences for pinning or floating (Figure~\ref{fig:versioning-strategy}).
We categorize a project's overall versioning strategy into \emph{floating only}, \emph{pinning only}, or \emph{mixed} based on the version constraints of its production dependencies.
Similar to the trend of version constraints, the popularity of \emph{floating only} strategy increased from 2011 to 2018 but has fallen since then, replaced by more adoption of \emph{pinning only} and \emph{mixed} strategy in the last five years.
Our results indicate that although projects sometimes make deliberate decisions per dependency~\cite{DBLP:journals/tosem/JafariCSA23}, the majority of projects make a uniform decision, motivating us to use similar alternations in our simulations (Section~\ref{sec:method-rq1}).

\subsection{Method}
\label{sec:method-rq1}

\paragraph{Overview}
At a high level, we take a \emph{counterfactual analysis} approach~\cite{morgan2015counterfactuals}---that is, estimating the causal effect of pinning by comparing the ``what-if'' differences with not pinning in each project.
This leads to our design of comparing metrics 
between a \emph{control} condition consisting of our observed projects, and a (simulated) \emph{treatment} condition wherein we alter each of the project's \Code{package.json} to pin all its direct dependencies.
We design a set of five outcome metrics in the project dependency graph to measure the five trade-offs raised in favor or against pinning, previously summarized in Table~\ref{tab:trade-offs}.
However, the comparison of these metrics is challenging as we would need to simulate how pinning would change the dependency graph of a focal project.
What's more, the values of metrics may change over time as the focal project and the ecosystem continuously evolves.
To overcome the first challenge, we leverage a hack in the npm dependency resolver for accurate time-traveling dependency resolution~\cite{DBLP:conf/msr/PinckneyCGB23}.
To overcome the second challenge, we organize the simulated data into \textit{balanced panels}, wherein each project is observed and simulated at five different times under both conditions (control and treatment). 
Then, we use \emph{fixed-effects panel regression} to estimate the effect of pinning 
separately for each of the five outcome metrics. 
Panel regression is a statistically sound econometric technique to model double-dimensional data (data collected from multiple units across different time periods) robustly to unobserved heterogeneity (i.e., project/time-specific characteristics that affect the outcome metric but are not directly measured or included in the model)~\cite{bruderl2015fixed, hanck2021introduction}.
Next, we give more details on the simulation setup and regression analysis.

\paragraph{Simulation Setup}
To implement the above design, we get the latest revision between September 2021 and September 2022 for each project in our dataset.
Then, we use September 12, 2022 as a starting time point for simulations (denoted $t_0$). 
We recreate the state of each project at this timestamp $t_0$, including the \Code{package.json} files, and resolve the corresponding dependency graphs (denoted as $G_{t0}$). 
Such historical resolution is non-trivial as we will need to reconstruct the state of the npm ecosystem at those times. 
Fortunately, thanks to a previous report by \citet{DBLP:conf/msr/PinckneyCGB23}, we use the undocumented \Code{\,-{}-before\,} argument in the npm command-line tool to enable \textit{time-traveling} dependency resolution.
Next, we create an altered version of each \Code{package.json} file by removing the ``caret'' and ``tilde'', i.e., pinning all direct dependencies to the minimal version specified in \Code{package.json}, and resolve the simulated dependency graph ($G'_{t0}$) analogously. 
This step results in two observations for each project, that feed into our ``control'' and ``treatment'' groups, respectively.
Similarly, we check out four additional versions of the \Code{package.json} file, at four evenly-spaced fixed 90-day intervals from December 11, 2022 to September 7, 2023 (denoted $t_1 \ldots t_4$). At each time point, we resolve the corresponding original and simulated (all-pinning) dependency graphs, which adds four additional observations per project into each of our ``control''  and ``treatment'' groups.
Note that for each project and treatment condition, we intentionally resolve dependency graphs at different time points from the same \Code{package.json} file; this setting, combined with our outcome metrics below, provides a framework to estimate the costs of action (or risks of not taking action) over the observation period, independent of the actual project actions taken during that period.

\paragraph{Outcome Metrics}
For each of the five original and simulated dependency graphs resolved for a given project, we compute five security- and maintenance-related metrics, each corresponding to one dimension of trade-offs in Table~\ref{tab:trade-offs}.
Their definitions are as follows:
\begin{enumerate}[leftmargin=20pt]
    \item \textsf{\# of floating dependencies} (\Code{n\_floating}):  The number of edges in the dependency graph with a floating version constraint. This metric captures \textit{attack surface}, i.e., the number of packages where a malicious update would be installed automatically if they were attacked.
    \item \textsf{\# of automatic updates} (\Code{n\_auto\_updates}):  The number of automatic updates happened between $t_0$ and $t$ for the dependency graph at time $t$. This metric captures the risk of \emph{breaking changes}, assuming that breaking changes happen uniformly randomly among new releases.
    \item \textsf{\# of security vulnerabilities} (\Code{n\_vuln}): The number of known vulnerable dependencies in the dependency graph based on entries in GitHub Advisory~\cite{github-adv} at time $t$. This metric captures the maintenance cost introduced by the need to remove \emph{security vulnerabilities} in the project (usually by manually updating to a newer version). 
    \item \textsf{\# of outdated dependencies} (\Code{n\_outdated\_deps}):  The number of outdated direct dependencies in the dependency graph is based on whether a newer release exists at time $t$. This metric captures the maintenance cost introduced by the need to manually update \emph{outdated dependencies}. 
    \item \textsf{\# of bloated dependencies} (\Code{n\_bloated}):   The number of packages with more than one version simultaneously present in the dependency graph (i.e., bloated). This metric captures the maintenance cost introduced by \emph{dependency conflicts} because npm resolves such conflicts by allowing different versions of the same package at different levels~\cite{how-npm3-works}.
\end{enumerate}
Intuitively,  \Code{n\_floating} and \Code{n\_auto\_updates} should decrease with pinning and increase with floating;
\Code{n\_vuln}, \Code{n\_outdated\_deps} and \Code{n\_bloated} should decrease with floating and increase with pinning. We designed all metrics such that lower values are generally better (e.g., lower costs or risk).

\paragraph{Panel Regression}
For each metric $M$, we fit the following panel regression model: %
\begin{equation}
\label{eq:plm}
\ln(M+1) \sim \text{\emph{pinning}} + \ln(size(G)) + \text{\emph{pinning}} \times \ln(size(G)) + \alpha + \beta 
\end{equation}
In this model specification, $\alpha$ and $\beta$ represent project-fixed and time-fixed effects, respectively. 
These two variables give the model robustness to unobserved heterogeneity, allowing us to model the unmeasured project-specific or time-specific characteristics that might affect the dependent variable.
The main variable of interest is the \textit{pinning} flag, whose estimated coefficient, if statistically significantly different from zero, represents the average change in metric $M$ relative to the default ``control'' condition, while holding fixed the other factors.
Finally, we model an explicit interaction with the size of the dependency graph $size(G)$ (i.e., number of nodes), to test whether the effect of pinning varies with $size(G)$---one can expect, e.g., that having more direct and transitive dependencies increases the attack surface of a downstream project.
The models are fitted using the R \Code{plm} package~\cite{r-plm}.

\paragraph{Limitations and Threats to Validity}

Our model design should allow reliable estimates of the impact of pinning, but it relies on accurate dependency resolution at past points in time.
While we believe that our solution is accurate, simulations can still deviate from the realities in the past due to npm behavior changes, configuration differences, etc. 
The dependency resolution attempts may also fail with an error, upon which we discard the data point: In total, we have a success ratio of 83.6\% for npm packages and 87.7\% for GitHub repositories.
For the failed resolutions, npm's error logs indicate the failures can be due to \emph{bad version constraints} (41.9\%), \emph{missing packages} (37.0\%), \emph{git errors} (8.9\%), \emph{bad URLs} (5.7\%), \emph{missing local files} (2.6\%), \emph{incompatible OS} (2.4\%), and others.
Resolution failures can happen for many reasons, such as a release being taken down by npm and a project depending on a package not present in the npm registry (i.e., likely proprietary).
As such software builds are inherently volatile~\cite{DBLP:conf/icse/SeoSEAB14, DBLP:conf/icse/DmeiriTWBLDVR19}, we believe these errors are difficult, if not impossible, to fix in an automated simulation pipeline.
Thus, we conclude that the success ratio is acceptable and the scale of our data is sufficient for our study design, being comparable with or larger than datasets in the medical and econometrics literature where similar designs are adopted (e.g.,~\cite{crouse1999randomized, stutzer2006does, gerstorf2010late}).

\setlength{\abovecaptionskip}{2pt}%
\setlength{\intextsep}{0pt}
\begin{wrapfigure}{r}{0.4\textwidth}
    \includegraphics[width=\linewidth]{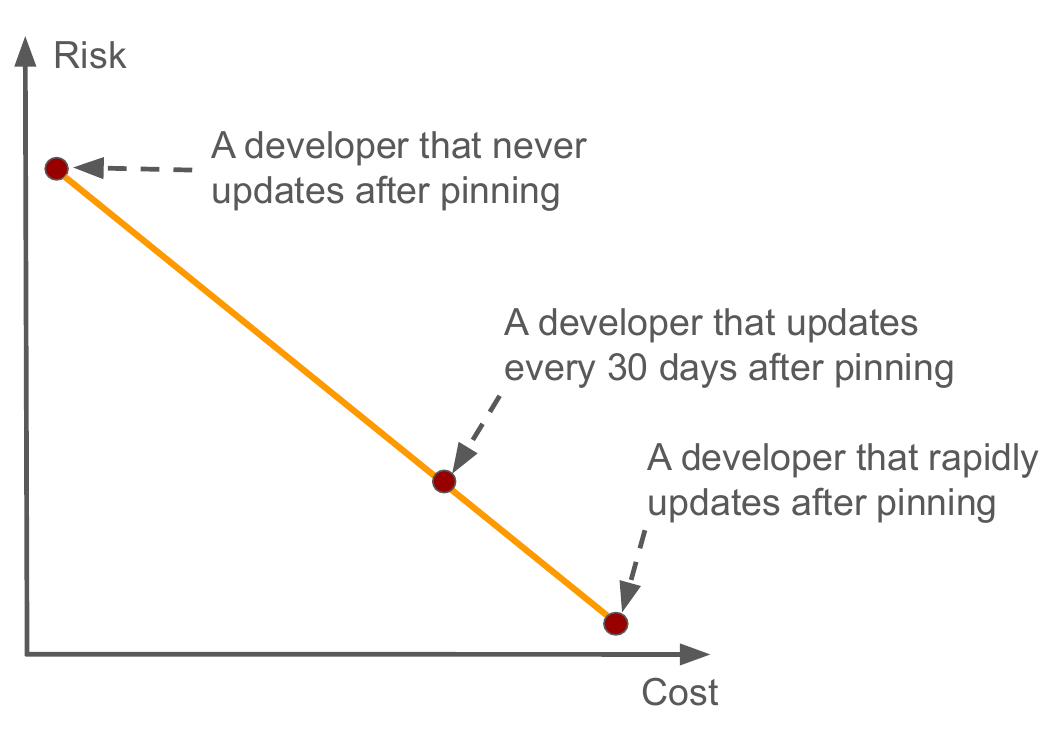}
    \caption{An illustration of the trade-off between cost and risk after pinning.}
    \label{fig:tradeoff-pinning}
\end{wrapfigure}
\setlength{\abovecaptionskip}{10pt}

Our simulation setup enables a holistic approach to measure the costs of taking action (or the risks of not), but does not fully take into account the complexity of project actions that may happen over a year.
For example, a project that opts for pinning can invest no effort in updating its pinned dependencies, rapidly update upon every new release of a dependency, or take an intermediate approach (e.g., update every 30 days).
In other words, for at least some of our outcome metrics (e.g., security vulnerabilities), developers using a pinning strategy can choose how much effort to invest in manual actions to further trade-off costs and risks (Figure~\ref{fig:tradeoff-pinning}).
In this paper, all outcome metrics are estimated from the conservative case of no manual developer actions; we leave the exploration of other cases to future work.

In terms of the metric design, we implicitly assume that malicious package updates and breaking changes follow a uniform random distribution, counting potential rather than actual events.
For malicious package updates, real attackers are generally opportunistic, and existing research does not show any noticeable statistical patterns in attack target selection~\cite{DBLP:conf/dimva/OhmPS020, DBLP:conf/sp/LadisaPMB23}.
Thus, we believe it is reasonable to assume attacks to happen uniformly randomly among packages.
For breaking changes, it could be possible to get more accurate estimations from static analysis or regression testing, but adopting either would severely limit and bias the amount of data available for modeling purposes.
Specifically, (1) static analysis only works well on TypeScript packages, (2) regression testing is only applicable to projects with a running test suite and prior regression testing studies collected at most a few hundred JavaScript projects with breaking changes~\cite{DBLP:conf/ecoop/MezzettiMT18, kong2024towards}.
In fact, we believe that breaking changes are so hard to avoid---even in well-managed projects (see Hyrum's Law~\cite{hyrum-law})---and well-tolerated in the npm ecosystem~\cite{DBLP:journals/tosem/BogartKHT21} that the uniformly random assumption is actually a reasonable approximation to quantify the expected risk (or cost) of breaking changes on a large scale npm based dataset.

In terms of external validity, it is important to note that our study are specific to the npm ecosystem and the results may not generalize to other ecosystems that adopt different baseline practices.
For example, the Maven ecosystem (Java) is smaller~\cite{module-counts}, conducts pinning extensively~\cite{DBLP:conf/msr/0001PSTB19}, and handles dependency conflicts differently~\cite{DBLP:conf/sigsoft/WangWLWWYYZC18}, so the trade-off between pinning and floating could manifest in a different manner.
In addition, generalizations to closed-source applications should also be careful as they may have different characteristics compared with our dataset. 

\subsection{Results}
\label{sec:results-rq2}

In this section, we report results based on \emph{production dependency graphs} (i.e., excluding development dependencies) from all projects in all time points (Table~\ref{tab:regression}).
Wald tests show that the full model (with pinning and the interaction) is preferable for all five outcome variables ($p<0.001$) compared to a null model with only $size(G)$ as an independent variable.
Note while development dependencies are generally considered irrelevant for security vulnerabilities~\cite{DBLP:conf/kbse/LatendresseMCS22, DBLP:conf/esem/PashchenkoPPSM18}, they form an important attack surface for malicious package updates that compromise build systems~\cite{DBLP:conf/sp/LadisaPMB23, DBLP:conf/dimva/OhmPS020}.
Therefore, we have also explored the robustness of our findings: (1) in dependency graphs with development dependencies, (2) separately on npm packages and GitHub repositories, and (3) using only data collected at $t_0$. 
The results are highly consistent and available in the replication package (Section~
\ref{sec:data-availability}).

\begin{table}
    \footnotesize
    \renewcommand{\arraystretch}{1.2}
    \centering
    \caption{The descriptive statistics (aggregated over all time points) and the longitudinal evolution of the studied security and maintenance related metrics between the control and treatment group (pinning).
    }
    \begin{tabular}{llrrrrrrc}
    \toprule
        Metric & Condition & Min & 25\% & Median & 75\% & Max & Mean & Evolution  \\
    \midrule
       \texttt{n\_floating} & \includegraphics[height=1.5mm]{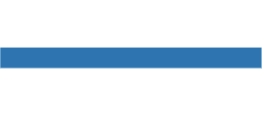} Control & 0 & 8 & 65 & 335 &  9,191 & 432.35 &\multirow{2}{*}{\adjustimage{height=10mm,valign=m}{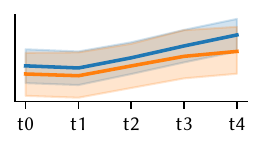}} \\
       & \includegraphics[height=1.5mm]{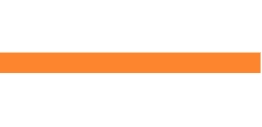} Pinning & 0 & 5 & 60 & 327 & 8,909 & 427.83 & \\\addlinespace
       \texttt{n\_auto\_updates} & \includegraphics[height=1.5mm]{figs/legend_control.png} Control  & 0 & 0 & 3 & 21 & 1,354 & 30.30 & \multirow{2}{*}{\adjustimage{height=10mm,valign=m}{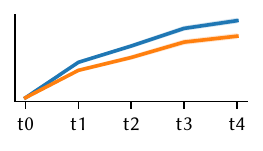}}\\
       & \includegraphics[height=1.5mm]{figs/legend_pinning.png} Pinning & 0 & 0 & 2 & 14 & 662 & 24.86  \\\addlinespace
       \texttt{n\_vuln} & \includegraphics[height=1.5mm]{figs/legend_control.png} Control & 0 & 0 & 0 & 2 & 85 & 2.13  & \multirow{2}{*}{\adjustimage{height=10mm,valign=m}{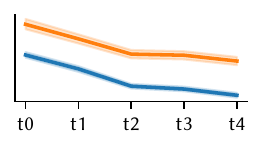}} \\
       & \includegraphics[height=1.5mm]{figs/legend_pinning.png} Pinning & 0 & 0 & 0 & 2& 97&  2.70  \\\addlinespace
       \texttt{n\_outdated\_deps} & \includegraphics[height=1.5mm]{figs/legend_control.png} Control & 0 & 1 & 2 & 7 & 225 & 5.96  & \multirow{2}{*}{\adjustimage{height=10mm,valign=m}{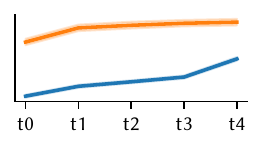}} \\
       & \includegraphics[height=1.5mm]{figs/legend_pinning.png} Pinning & 0 & 1 & 4 & 11 & 302 & 8.78  \\\addlinespace
       \texttt{n\_bloated} & \includegraphics[height=1.5mm]{figs/legend_control.png} Control & 0 & 0 & 1 & 13 & 650 &  22.38  & \multirow{2}{*}{\adjustimage{height=10mm,valign=m}{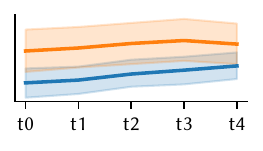}}  \\
       & \includegraphics[height=1.5mm]{figs/legend_pinning.png} Pinning & 0 & 0 & 1 & 15& 639 & 23.51  \\\addlinespace
    \bottomrule
    \end{tabular}
    \label{tab:metric-values}
    \renewcommand{\arraystretch}{1}
\end{table}

\begin{table}
\footnotesize
\centering 
\caption{The fixed-effect panel regression results. 
We obtain a high goodness-of-fit for \texttt{n\_floating} (R$^{2}$$\ge$0.6), acceptable goodness-of-fit for \texttt{n\_vuln}, \texttt{n\_outdated\_deps}, and \texttt{n\_bloated} (R$^{2}$$\ge$0.1), and low goodness-of-fit for \texttt{n\_auto\_updates} (R$^{2}$$<$0.1).
The R$^{2}$ values are comparable with similar prior studies~\cite[see, e.g.,][]{DBLP:conf/sigsoft/ValievVH18, DBLP:conf/icse/ZahanSHW23}. 
Note that since our goal is explanatory (not predictive), we do not evaluate models based on on R$^{2}$ values, but conduct a series of robustness checks to see whether the findings are robust over different settings (Section~\ref{sec:results-rq2}).
We report local effect sizes computed using Cohen's $f^2$~\cite{selya2012practical}.
}
\label{tab:regression} 
\begin{tabular}{lrrrrr} 
  \toprule
   & \multicolumn{5}{c}{$\ln($\emph{Dependent Variable}$) + 1$} \\ 
   & \texttt{n\_floating} & \texttt{n\_auto\_updates} & \texttt{n\_vuln} & \texttt{n\_outdated\_deps} & \texttt{n\_bloated}\\
  \midrule
  \emph{pinning} & $-$0.519$^{***}_{+++}$ & $-$0.274$^{***}_{+}$ & 0.011$^{***}_{+}$ & 0.302$^{***}_{+++}$ & 0.029$^{***}_{+}$\\ 
   & (0.006) & (0.005) & (0.005) & (0.005) & (0.003)\\ \addlinespace
  $\ln(size(G))$ & 1.084$^{***}_{+++}$ & 0.794$^{***}_{+}$ & 0.283$^{***}_{+}$ & 0.068$^{***}$ & 0.631$^{***}_{++}$ \\ 
   & (0.014) & (0.036) & (0.021) & (0.011) & (0.027)\\\addlinespace 
  \emph{pinning} $\times \ln(size(G))$ & 0.084$^{***}_{+++}$ & 0.003 & 0.028$^{***}_{+}$ & 0.019$^{***}$ & 0.007$^{***}$\\
  & (0.001) & (0.001) & (0.001) & (0.001) & (0.001) \\
 \midrule
 R$^{2}$ & 0.715 & 0.053 & 0.154 & 0.330 & 0.227 \\ 
 Adjusted R$^{2}$ & 0.684 & $-$0.052 & 0.059 & 0.255 & 0.141\\ 
 \bottomrule
 \multicolumn{6}{r}{$^{***}$p$<$0.001, $_{+++}$ large effect size, $_{++}$ medium effect size,   $_{+}$ small effect size} \\ 
\end{tabular} 
\end{table} 

For the effect of pinning, we find that pinning all direct dependencies has a significant influence on the three measures of cost---higher exposure to security vulnerabilities (\Code{n\_vuln}), number of outdated dependencies that need to be manually updated (\Code{n\_outdated\_deps}), and higher exposure to dependency conflicts (\Code{n\_bloated}). Those results are supported by the differences in the descriptive results over different time periods (Table~\ref{tab:metric-values}) and the significant coefficients in our panel regression models ( Table~\ref{tab:regression}); the interaction term shows that all three costs of pinning increase with dependency graph sizes; the exposure to security vulnerabilities has the strongest interaction effect.

\begin{figure}[t]
    \centering
    \includegraphics[width=0.9\linewidth]{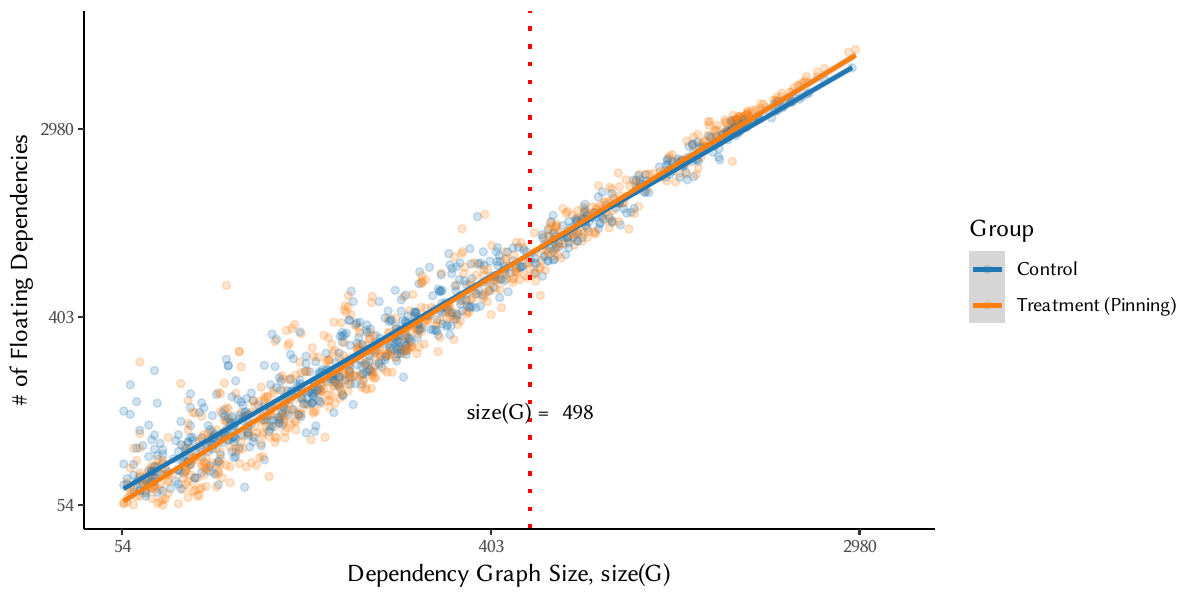}
    \caption{The effect of pinning direct dependencies on malicious package update attack surface (as measured by the number of floating dependencies in the entire dependency graph). Our model shows that pinning reduces attack surface for small dependency graphs, but the positive effect decreases as the dependency graphs grow larger and even flips to negative for dependency graphs with more than 498 dependencies.
    }
    \label{fig:n-floating}
\end{figure}

The effect of pinning on the two metrics expected to show benefits of pinning (reduced attack surface and fewer breaking changes) is dim.
For the number of automatic updates, the regression model fits poorly and provides little explanatory power (other factors like picking dependencies with fewer releases likely have a stronger effect than pinning). 
For the number of floating dependencies, the interaction effect with the size of the dependency graph (see Table~\ref{tab:regression} and illustration in Figure~\ref{fig:n-floating}) shows that the benefit of pinning on reducing the attack surface is smaller with larger dependency graphs and even (counter-intuitively) negative for very large dependency graphs.

\section{RQ2:  How effective is coordinating pinning among selected upstream packages in reducing the risk of malicious package updates ecosystem-wide? }

As the results from RQ1  (Section~\ref{sec:results-rq2}) indicate, pinning direct dependencies is ineffective and possibly even harmful for preventing malicious package updates if a project has a large and complex dependency graph (we will return to this issue in Section~\ref{sec:discussion}). 
Then, a natural follow-up question arises: Are there better, alternative interventions that can effectively reduce the risk of malicious package updates in the npm ecosystem?
Inspired by previous research that reveals the structures of dependency networks~\cite{DBLP:journals/ese/DecanMG19, DBLP:conf/uss/ZimmermannSTP19}, we hypothesize that \emph{the risk of malicious package updates can be greatly reduced if a small set of ``core'' packages pin and promptly update all their dependencies after proper auditing} (ensuring there is no outdated dependency but also no malicious update).
That is, instead of all developers redundantly performing security actions, a few package maintainers (or a dedicated security team for the ecosystem) might effectively take on the cost of (some) security work by pinning and actively managing dependencies, so that all developers benefit. This would enable a vision where pooling some community resources would be much more efficient and effective to achieve security outcomes than trying to convince every single maintainer to change their practices (e.g., in a way similar to how the Linux Foundation starts the Core Infrastructure
Initiative in response to HeartBleed~\cite{DBLP:conf/msr/Walden20} or how Google provides the Assured Open Source Software service~\cite{assured-open-source}).
This leads to our RQ2 of exploring how coordinated pinning in select upstream packages may reduce the overall risk of malicious package updates at an ecosystem level.

\subsection{Method}
\label{sec:method-rq3}

\paragraph{Overview}
In a nutshell, we quantify the effect a few maintainers can have on 
the risk of (simulated) malicious package updates on the entire npm ecosystem.
We assume that we have the ability and budget to convince $n$ package maintainers of our choice to change their dependency management practices as follows: They pin each of their dependencies, review all dependency updates to catch malicious updates, and rapidly update all non-malicious updates.
This way, we would expect that some malicious updates are effectively detected by these $n$ maintainers, preventing them from affecting other developers in the ecosystem.
{Focusing on whether this approach is feasible at all, we make the following optimistic, simplifying assumptions: The $n$ maintainers review and update all dependencies immediately (without adding latency)
and they recognize all malicious package updates when they audit an update. 
We also assume that reporting and taking down a malicious package takes some time (\citet{DBLP:conf/dimva/OhmPS020} report a mean of 209 days), so even if some maintainers detect a malicious update and protect their users, others may still be exposed.
} 
Finally, we assume all other developers are floating all their dependencies, as they already mostly do. 

\paragraph{Dependency Network}
We simulate attacks on a snapshot of the package-level npm dependency network constructed from dependency relationships specified in the latest version of each package (cut-off in September 2023), using the same npm-follower dataset~\cite{DBLP:conf/sigsoft/PinckneyCG023} in Section~\ref{sec:data-collection}. 
The network has 1,566,434 packages (nodes) and 7,873,239 dependency edges.

\paragraph{Attack Selection Strategy}
In practice, attackers might target specific applications (e.g., specifically attacking cryptocurrency packages~\cite{crypto-attack}) or they may try to compromise whatever package they can to opportunistically cause damage (e.g., as common for untargeted ransomware attacks). Since we do not know what exact strategy attackers may pursue and since their strategy may change over time, we make the simplifying assumption that attackers will try to compromise high-impact targets by (randomly) selecting any of the top-$m$ packages (denoted as $A$) on npm as ranked by the downloads adjusted impact defined in Equation~\ref{eq:impact} (i.e., impact-based strategy).

\begin{figure}[t]
    \centering
    \subfigure[\textbf{Attack Impact Computation:} Packages C to H all directly or indirectly depend on B and are hence reachable if B is attacked, but package A is not. The impact of compromising package B is the sum of downloads of all reachable packages (C--H) divided by the total number of downloads across all packages. In this case, $\textit{impact}(B) = (1m+10k+5k+2k+1k+900+500) / (90k+1m+10k+5k+2k+1k+900+500) \approx 0.92$. ]
    {\includegraphics[width=0.48\linewidth]{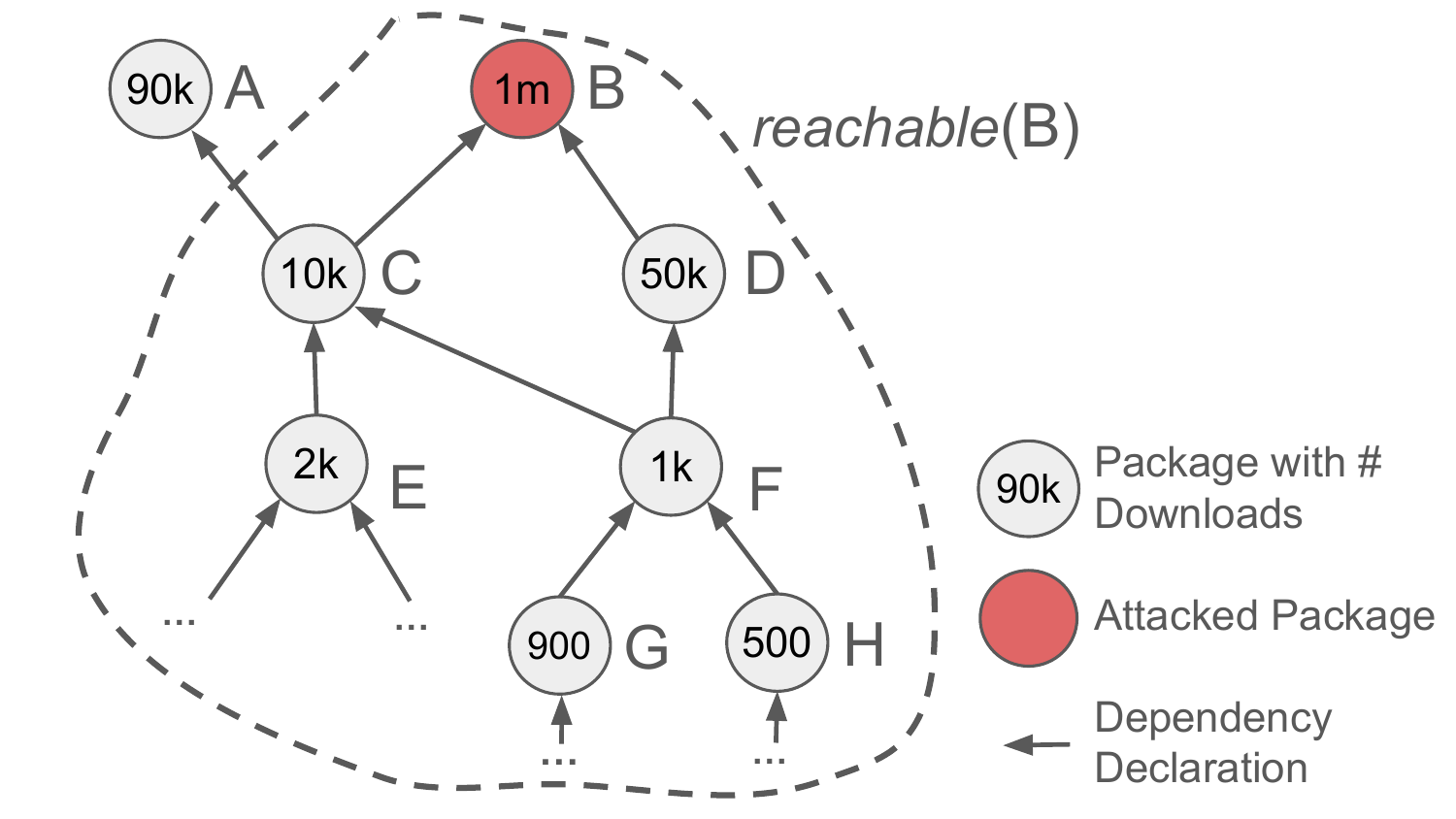}\label{fig:example-reachability}}
    \hfill
    \subfigure[\textbf{Local/Transitive Pinning:} Assume B is being attacked.
    If C pins its dependencies, the attack would not reach E, as C would not update to the malicious version. 
    However, local pinning of F would have no effect, since the attack still propagates through the floating edges originating from C and D. In contrast, transitive pinning of F would protect G and H (and F's other dependents) from the attack.]{\includegraphics[width=0.48\linewidth]{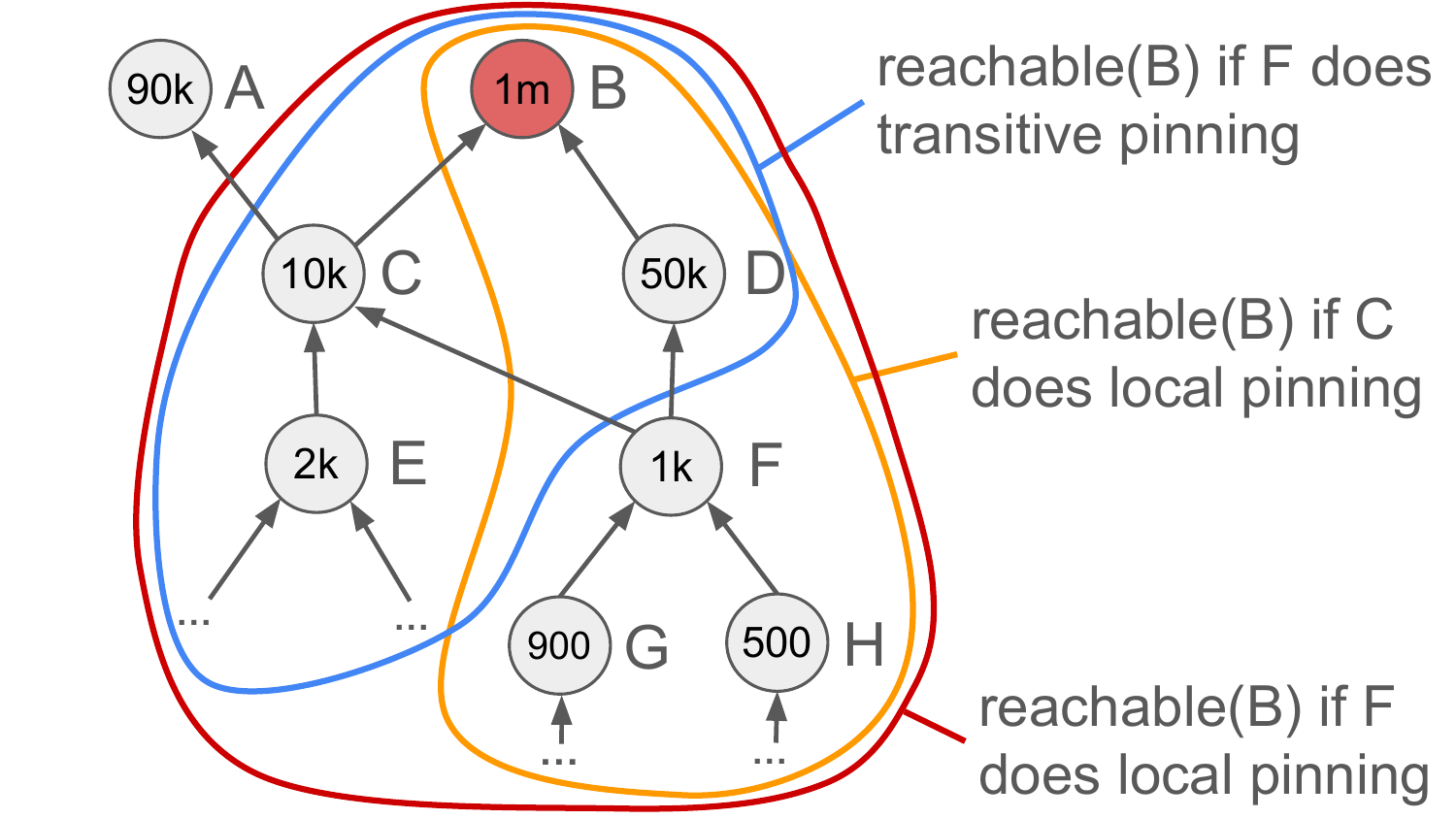}\label{fig:example-transitive-pinning}}
    \vspace{-1mm}
    \caption{Examples to illustrate attack impact computation and how local/transitive pinning works.}
    \vspace{-1mm}
    \label{fig:example-rq3}
\end{figure}

\paragraph{Risk Metric}
Given the set $P$ of all packages in the network, we quantify the impact of compromising a package $a\in P$ as the download-weighted coverage of its reachable dependents within the entire npm network, where every package that depends directly or indirectly on package $a$ is considered reachable of the edge to $a$ in its dependency graph is floating:
\begin{equation}
\label{eq:impact}
    \textit{impact}(a)=\frac{\sum_{p \in \textit{reachable}(a)}\textit{downloads}(p)}{\sum_{p \in P}{\textit{downloads}(p)}}
\end{equation}
In other words, $\textit{impact}(a) \in [0, 1]$ measures the fraction of npm downloads that will be affected if package $a$ is compromised, as illustrated in Figure~\ref{fig:example-reachability}.

For the set of packages $A\subset P$ (i.e., the top-$m$ most impactful packages), in which an attacker may target any of $p \in A$ based on the attack selection strategy, we define \textit{risk(A)} as the arithmetic mean of the impact across all possible targets in $A$ an attacker can choose from (i.e., $\textstyle\textit{risk}(A)=\frac{1}{{|A|}}{\sum_{a \in A} \textit{impact}(a)}$).
Intuitively, $risk(A) \in [0, 1]$ is an estimation of the expected average percentage of npm downloads that would be affected by compromising any one of the packages in $A$.  

\paragraph{Pinning Mechanism}
With npm's existing mechanisms, the maintainers of the $n$  defending packages can only pin their direct dependencies (\textit{local pinning}), but not their transitive ones. Maintainers can freeze their dependency graph with a \Code{package-lock.json} file, which, with npm's current design, only has an effect on their own package and not on anyone importing their package.

To give maintainers more power (at a higher maintenance cost), we imagine an alternative strategy of \textit{transitive pinning}, where maintainers can pin direct and indirect dependencies. This could be implemented with a 
change to npm that allows (a)~package maintainers to upload their \Code{package-lock.json} as part of the deployed package and (b)~npm to consider that file in each package during dependency resolution. We illustrate the difference in Figure~\ref{fig:example-transitive-pinning}.

\paragraph{Defense Selection Strategy}
Finally, we need to decide which $n$ packages to defend with local or transitive pinning.
Unfortunately, the problem of choosing an optimal package set of size $n$ that minimizes $risk(A)$ is equivalent to the Influence Minimization problem, which is NP-hard~\cite{DBLP:conf/icde/0002ZW0023}.
Hence, we focus on evaluating simple heuristics to choose the defended packages and leave the optimization as future work.
Specifically, we choose top-$n$ packages with (1) highest downloads, (2) highest out degrees (i.e., the packages with the most \textit{direct} dependencies), (3) highest betweenness centrality (i.e., the packages on most dependency \textit{paths}), (4) highest downloads $\times$ out degrees (i.e., popular packages with many direct dependencies), and (5) highest downloads $\times$ betweenness centrality (i.e., popular packages on many dependency paths).
Intuitively, we expect defense selection strategies based on out degree to work best for local pinning as they emphasize direct dependencies, and strategies based on betweenness to work best for transitive pinning as transitive pinning interrupts paths.

\paragraph{Simulation}
For each combination of pinning mechanism, defense selection strategy, and number of defended packages from 0 to $n$, we measure the impact of the defense on the risk metric. To compute the impact, we compute \textit{risk(A)} with a modified \textit{reachable} function (Equation~\ref{eq:impact}, different from the standard notion of graph reachability) by breaking propagation of the attack when the final edge is pinned (local pinning) or if the path contains a defended package (transitive pinning).

\paragraph{Limitations and Threats to Validity}
To make the simulation feasible, we make several simplifying assumptions described above.
Notably, we assume that  attacks happen uniformly randomly on a fixed number of attack targets, while real attack selection strategies may be more targeted, more opportunistic, and may adapt in response to defense strategies.
Therefore, we additionally explored two different attack selection strategies: (1) attacking any actively maintained projects randomly, and (2) attacking the most fragile or under-maintained packages (e.g., as the \Code{xz} attack, which focused on a dependency with a single overworked maintainer~\cite{xz-utils}) defined as the bottom-1000 packages in our package dataset, ranked by the number of core maintainers estimated in World of Code~\cite{DBLP:journals/ese/MaDBAVTKZM21}. For space and since our results are generally robust under the different attack selection strategies, we only report results under the impact-based attack selection strategy (the remaining are in the replication package, Section~\ref{sec:data-availability}).
What's more, the dependency network is only an approximation of reverse dependency relationships. 
However, similar kinds of dependency networks are frequently used in prior research~\cite{DBLP:journals/ese/DecanMG19, DBLP:conf/uss/ZimmermannSTP19} because it is generally impossible to compute precise reverse dependencies as would be resolved by npm due to its deduplication optimizations (Section~\ref{sec:background}).
For defense selection strategies, we have to rely on simple heuristics, leaving potentially better strategies (e.g., more computationally expensive optimization attempts) to future work.
Finally, due to a lack of access to real-world application data, our simulation does not accurately quantify the impact of malicious package updates on applications, whether open-source repositories or closed-source projects outside of npm.
We mitigate this by considering downloads in attack impact estimation (Equation~\ref{eq:impact}), but npm download data can be also noisy and inconsistent.
Due to these simplifying assumptions, readers should take our results as a first exploration of broad defense strategies and their ecosystem-wide effects, rather than specific implementation recommendations.

\subsection{Results}
\label{sec:results-rq3}

\begin{figure}[t]
    \centering
    \includegraphics[width=0.98\linewidth]{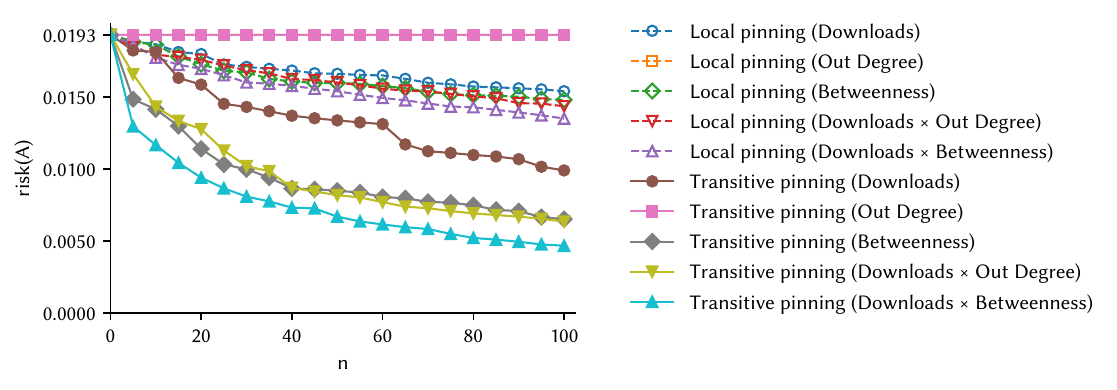}
    \vspace{-3mm}
    \caption{The reduction of malicious package update risk ($risk(A)$) by securing top $n$ packages with five different defense selection strategies using local pinning and transitive pinning, respectively.}
    \label{fig:intervention}
\end{figure}

In this section, we report results for $n = 100$ and $m=1,000$ (i.e., defending up to 100 packages to minimize the risk of attack among the top 1000 most impactful packages; the results are robust under other $n$, $m$ variations which are available in the replication package, Section~\ref{sec:data-availability}).
As visible in Figure~\ref{fig:intervention}, one would expect 1.93\% of npm downloads to be affected on average upon an attack without any intervention, which is enormous considering that npm has  $\sim$12 billion downloads per day in 2024~\cite{sonatype-report}.
Most defense strategies can reduce ecosystem-wide risk (from  doing nothing at $n=0$) with targeted defenses at 10 to 100 packages, but their effectiveness varies widely. The best local pinning strategy can reduce ecosystem-wide risk against the 1000 most popular packages by 12.21\% with defending only 20 packages and 30.09\% when defending 100 packages. 
The best transitive pinning strategy can reduce ecosystem-wide risk against the 1000 most popular packages by 51.34\% when defending only 20 packages and 75.81\% when defending 100 packages.
For both pinning mechanisms, selecting packages with betweenness weighted by downloads is most effective,
but transitive pinning is usually two to three times more effective than local pinning. 

\section{Discussion}
\label{sec:discussion}

\subsection{Why can pinning direct dependencies lead to a larger total number of floating dependencies (i.e., a larger attack surface for malicious package updates)?}

\begin{figure}[t]
    \centering
    \subfigure[Not pinning (\textbf{5} floating edges)]{\includegraphics[width=0.46\linewidth]{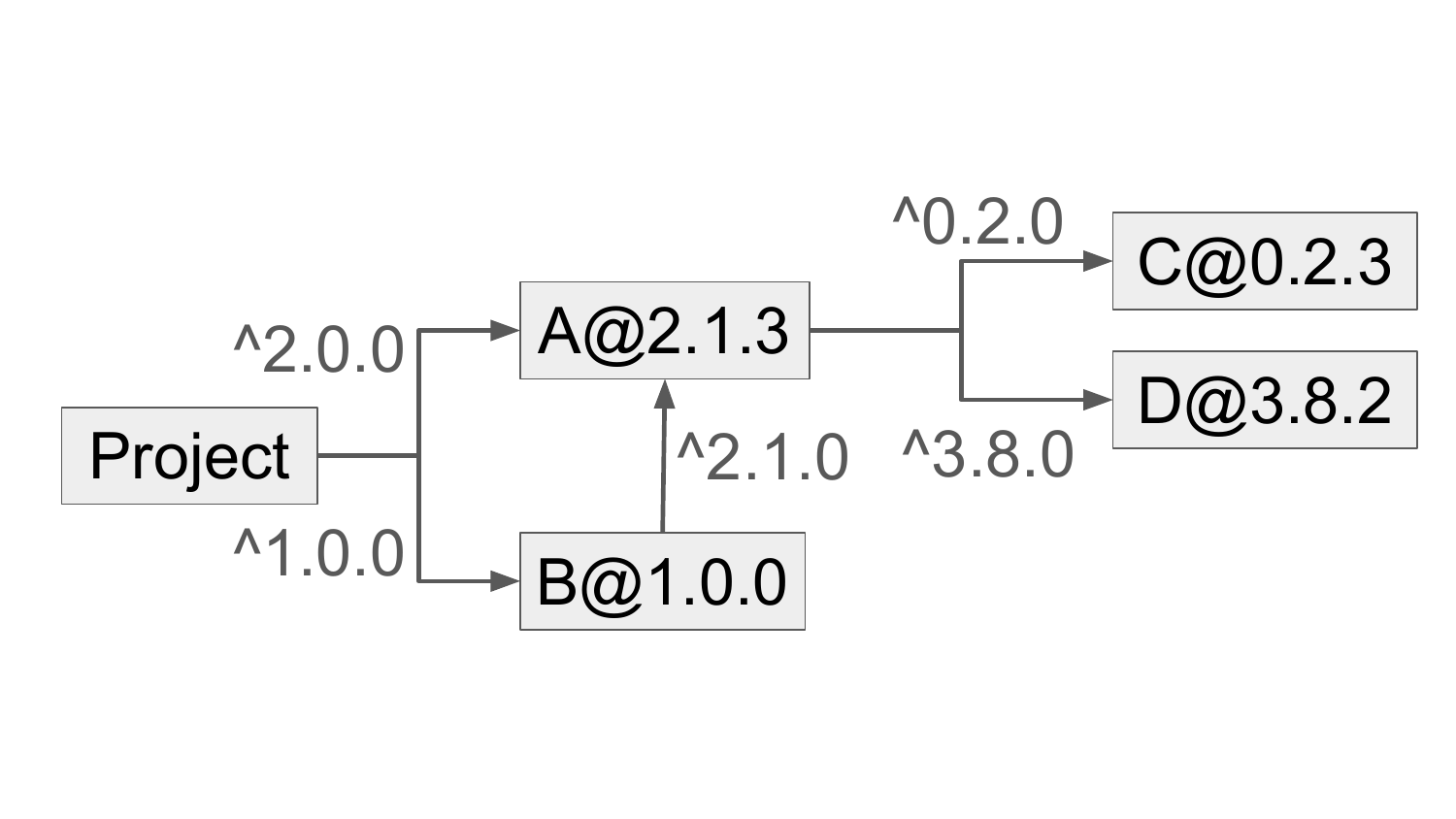}\label{fig:floating-example}}
    \subfigure[Pinning (\textbf{6} floating edges)]{\includegraphics[width=0.46\linewidth]{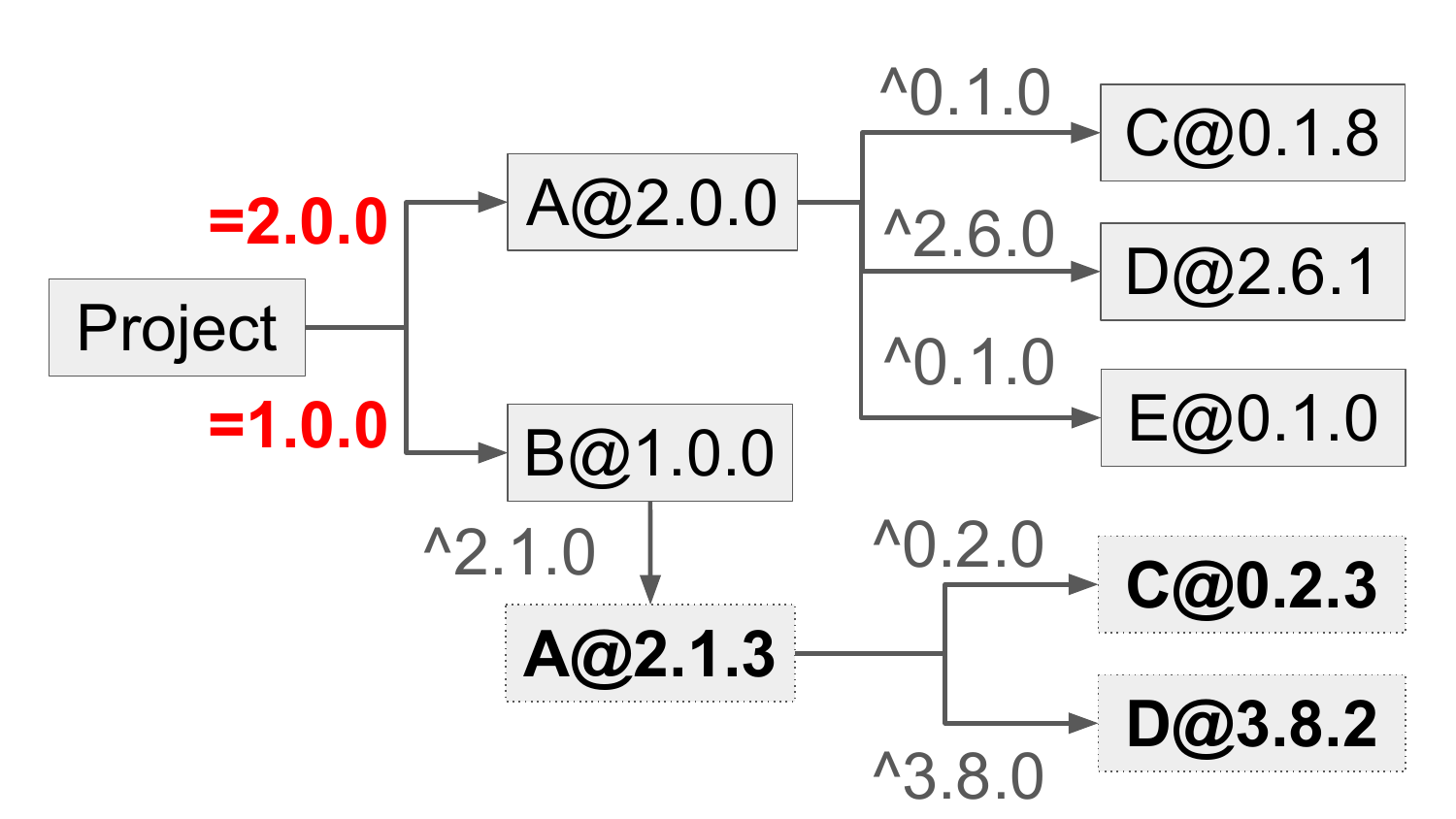}\label{fig:pinning-example}}
    \vspace{-1mm}
    \caption{A minimal example illustrating how pinning direct dependencies can cause dependency conflicts and lead to a bloated dependency graph with more transitive floating dependencies. Note that this is only an artificial example for simplicity; real-world cases happen mostly in much larger dependency graphs.
    }
    \vspace{-1mm}
    \label{fig:rq2-example}
\end{figure}

\label{dedup}
This surprising effect is explained by a detail of how the npm dependency resolver handles dependency conflicts~\cite{how-npm3-works} (cf. Section~\ref{sec:background}):
If two or more nodes in the dependency graph declare a dependency  on the same package, npm will first check (with heuristics) whether there is a version of that package that meets all declared version constraints (as in our example in Figure~\ref{fig:floating-example}). If there are conflicting version constraints in dependency declarations, npm will install multiple versions of the package to satisfy all version constraints (as in our example in Figure~\ref{fig:pinning-example}). This deduplication (or debloating) process is designed to download fewer, often redundant, copies of packages in the installation process. In fact, there are many possible deduplication heuristics and researchers have explored how to improve this in various ways (e.g., using an SMT solver~\cite{DBLP:conf/icse/PinckneyCGBCG23}).

The deduplication process interacts with pinning: If more dependencies are pinned, the dependency graph resolution has fewer opportunities to deduplicate a specific package and will be more likely to create larger dependency graphs with multiple versions of the same package. Those different versions may also then include additional dependencies---as visible in our example in Figure~\ref{fig:rq2-example}. Both effects increase the attack surface, in that attackers have possibly more version ranges they can attack for the same packages and there could be even additional packages they can attack.
This leads to our conclusion that \textit{the practice of (local) pinning is futile and even counter-productive for the purpose of reducing floating dependencies (aka attack surface) in one's dependency graph}.

Note that the deduplication heuristics introduced here and the ability to install multiple versions of the same package are unique to npm~\cite{DBLP:journals/tosem/BogartKHT21}.
In most other dependency managers, the resolver must find a single version for each package and pinning can make this much harder (as captured also in our \Code{n\_bloated} metric for pinning).
In those ecosystems, pinning would arguably still have only marginal effects on attack surface reduction, but not negative ones.
However, the costs for resolving dependency conflicts would be much more substantial~\cite{DBLP:conf/sigsoft/WangWLWWYYZC18, DBLP:conf/icse/WangW0WLWYCX020}. 

\subsection{Implication for Practitioners}

\textbf{Dependency pinning is (as expected) an expensive intervention.}
Regarding the drawbacks of pinning, our RQ1 results align with the argued trade-offs in Table~\ref{tab:trade-offs}:
Pinning direct dependencies substantially increases the effort for developers to manually update (and possibly review) dependencies if they want to stay up to date, or face many outdated dependencies if they do not. 
Also, projects that pin their dependencies are, as expected, much more likely to be exposed to known vulnerabilities in their dependency graph if they do not constantly update to available patches. While bots and notification tools like \textit{Dependabot} can help with some of that work and even automatically run tests and merge updates (as covered in prior work~\cite[e.g.,][]{mirhosseini2017can,DBLP:conf/msr/AlfadelCSM21}), they also have their limitations, creating additional complexity (e.g., notification fatigue~\cite{DBLP:conf/sigsoft/HeHGZ21}).

\textbf{Pinning direct dependencies is futile and (surprisingly) can be even counter-productive as a security intervention if floating is the mainstream practice in the ecosystem.}
Our RQ1 results regarding the expected security benefits of pinning are surprising and counter-intuitive to practitioners (based on our experience when sharing the results).
The first reason is that in many projects, direct dependencies are far outnumbered by transitive dependencies for which version declarations are controlled by upstream packages (Table~\ref{tab:dataset}). 
Thus, the action of a single project has a relatively small impact on the total number of floating dependencies in its dependency graph, explaining the small effective size and the diminishing return of pinning as the dependency graph grow larger (as confirmed by the interaction term $\text{\emph{pinning}} \times \ln(size(G))$ in our panel regression model and illustrated visually in Figure~\ref{fig:n-floating}).
Importantly, the security benefit of pinning does not only diminish, but it \textit{reverses} and pinning becomes actually counterproductive in large dependency graphs ($\ge$498 nodes according to our model, also visible as a crossover point in Figure~\ref{fig:n-floating}), to the point that \textit{pinning direct dependencies leads to more floating dependencies in the entire dependency graph.}  
In other words, an attacker may be able to attack the same dependencies in more version ranges or may even be able to attack additional dependencies that are only (indirectly) included in some versions of some dependency.
More generally speaking, the perceived security trade-off between pinning (to reduce the risk of malicious updates) and floating (to get fewer security vulnerabilities) does not hold in practice if the upstream packages are mainly floating (e.g., in npm and PyPI). 

Following the discussion above, \textbf{in an ecosystem where floating is the mainstream practice, pinned dependencies do not necessarily signal that a project is more secure from malicious package updates, especially for projects with large dependency graphs.}
In such contexts, pinning can create substantial maintenance costs, for limited security benefits, if any. 
Therefore, we argue that pinning should not be considered when scoring the security practices of packages.

\textbf{Instead of pinning, application developers should freeze their entire dependency graph with a shared lock file.}
In the npm ecosystem, it means using and \emph{commiting} the \Code{package-lock.json} file (Section~\ref{sec:background}) to the version control repository (44.21\% of GitHub repositories in our dataset did this).
This practice enables the effective pinning of all direct and transitive dependencies and more importantly, prevents updates from being installed automatically in a new machine or in the CI/CD environment.
However, \emph{lock files are intended only for applications (not packages)}, and they are not designed to allow easy selective updates or audits of dependency updates---hence every developer would need to individually judge when to update which (possibly indirect) dependencies rather than delegating some of that judgment to package maintainers and ecosystem moderators.

\textbf{To secure an ecosystem from supply chain attacks, package developers may consider community-wide coordination from a collective pool of resources.}
For example, our RQ2 results indicate that collective pinning actions could indeed be fairly effective at reducing the risk of malicious package updates and \textit{localize} the cost to a few (20--100) carefully selected packages.
We believe it is much more achievable to convince (and raise labor or funds for) changing practices in 100 packages for the good of the entire ecosystem, than trying to convince almost every maintainer in npm individually to change their practices.
However, our results also show clearly that npm's current local pinning is far less effective than transitive pinning could be. 
While transitive pinning is far more expensive for the maintainer defending a package (i.e., they need to audit updates of transitive dependencies too, not just of direct ones), it also provides a bigger lever for security. 
Transitive pinning is also easier to explain to developers, whereas local pinning can provide a false sense of security with counter-intuitive consequences (as shown in our RQ1).
Therefore, \textbf{we suggest changes to npm infrastructure and tooling to enable transitive pinning (e.g., publishing a \Code{package-lock.json} file for downstream consumption).}

However, we must note that these kinds of interventions would probably still experience pushback and be costly to implement from such an established ecosystem (as of pushing any organizational change~\cite{schein2010organizational}).
To make the process easier, \textbf{we encourage foundations, security companies, or companies using open source to dedicate resources (labor, funds) to such efforts}.
Even if only a small number of maintainers opt to purse this (i.e., transitive pinning and extensively review their updates), it can have a large impact on the ecosystem.

Finally, it is worth noting that the whole debate around version constraints relies on one critical assumption: \emph{malicious package updates will go undetected in the ecosystem for a significant amount of time}, which is probably still the case as of now (as evidenced by the \Code{xz} attack~\cite{xz-utils}).
If malicious package updates are taken down faster or even stopped before being published in the package registry, their impact  will be much smaller.
Thus, \textbf{efforts are needed to make malicious package updates as short-lived as possible.}
We envision that if a small number of core ecosystem packages can pin and \emph{promptly review their updates}, they can also help in such early detection, for example, by leveraging the ongoing research about detecting malicious packages~\cite[e.g.,][]{DBLP:conf/icse/SejfiaS22, DBLP:journals/corr/abs-2403-12196}.

\subsection{Implication for Researchers}

Our study leaves a few open research questions.
First, our study only focuses on the npm ecosystem, and other ecosystems adopt different baseline dependency versioning practices~\cite{DBLP:journals/tosem/BogartKHT21, DBLP:conf/msr/0001PSTB19}.
Future research is necessary to explore the impact of pinning/floating under those variations.
Second, while prior research provides evidence that floating jeopardizes build stability~\cite[e.g.,][]{DBLP:conf/issta/MukherjeeAR21}, the \Code{n\_auto\_update} model in RQ1 only explains a small amount of variance in the data ($R^2$<0.1), so we expect future work to better establish the causal relationship between pinning and application stability.
Third, our RQ2 only takes a first step to explore possible ecosystem-level strategies, and we encourage more research in this direction. 
We suggest exploring better strategies than download-weighted betweenness and exploring how to consider users of packages outside npm.  
Researchers may also explore alternative simulation approaches going beyond our simplifying assumptions, and, for example, consider the specific existing choices of pinning and floating in the dependency graph when suggesting which packages to defend. 
Future research could also explore whether these strategies hold when baseline pinning and floating practices are different in other ecosystems~\cite{DBLP:journals/tosem/BogartKHT21, DBLP:conf/msr/0001PSTB19}.
Another remaining question is whether a maintainer's decision to pin their package's dependencies produces unintended non-local costs for package users. 
For example, it is possible that the pinning in some parts of the dependency graph can affect users negatively in terms of causing them dependency conflicts (cf. \Code{n\_bloated} in Section~\ref{sec:results-rq2}). 
Researchers may explore whether different defense selection strategies (e.g., explicitly coordinating dependency updates across packages) or explicitly advocating for floating outside of a few defended packages can reduce unintended dependency conflicts.
Again, we expect all these concerns to manifest differently in different ecosystems and we envision a broad range of future research to establish the security impact of versioning practices in a wide variety of contexts.

\section{Conclusion}

In this paper, we have presented a counterfactual analysis and simulation study on the impact of version constraints in the npm ecosystem.
We have explored the security and maintenance impact of local pinning direct dependencies and the effectiveness of pinning on ``important'' packages against possible malicious package updates.
Our study provides counter-evidence for the use of local pinning as a security best practice, but supports the effectiveness of coordinated pinning in critical open-source packages.
Our study leads to a series of implications for the design of dependency management tools and ecosystem-level coordination to enhance software supply chain security.

\subsubsection*{\textbf{Data Availability}}
\label{sec:data-availability}

We provide a replication package at \url{https://doi.org/10.5281/zenodo.14693233}.

\subsubsection*{\textbf{Acknowledgement}}

He's and Kästner's work was supported in part by the National Science Foundation (award 2206859).
Bogdan's work was supported in part by the National Science Foundation (award 2317168) and research awards from Google and the Digital Infrastructure Fund.
We would like to thank James Herbsleb, Narayan Ramasubbu, Rohan Padhye, and the participants of S3C2 quarterly meetings for their valuable feedback to the earlier versions of this work. 

\bibliographystyle{ACM-Reference-Format}
\bibliography{references}

\end{document}